\documentstyle[aps,prl,epsfig,twocolumn]{revtex}
\newcommand{\beeq}{\begin{equation}}
\newcommand{\eneq}{\end{equation}}
\newcommand{\beeqa}{\begin{eqnarray}}
\newcommand{\eneqa}{\end{eqnarray}}
\newcommand{\la}{\langle}
\newcommand{\ra}{\rangle}
\begin{document}
\twocolumn[\hsize\textwidth\columnwidth\hsize\csname
  @twocolumnfalse\endcsname
\title{
A New Relation between post and pre- optimal measurement states.}
\author{Chirag Dhara \dag } 
\address{St. Xavier's College, Mumbai-400 001, INDIA}
\author{N.D. Hari Dass \ddag} 
\address{Institute of Mathematical Sciences, Chennai-600 113, INDIA}
\maketitle
\begin{abstract}
When an optimal measurement $(S_{x}, S_{y}, S_{z})$ is made on a
qubit, and what we call an \emph{Mutually Unbiased Mixture - MUM} of 
the resulting ensembles is taken, then the post measurement density matrix is shown to be related to the 
pre-measurement density matrix through a simple {\em linear and universal} 
relation. It is
shown that such a relation holds {\em only} when the measurements are made
in \emph{Mutually Unbiased Bases - MUB} \cite{sch,woot,mubref}.
For Spin - 1/2 systems it is also shown explicitly that non - orthogonal 
measurements fail to give such a linear relation no matter how the ensembles 
are mixed. The result has been proved to be true for arbitrary quantum 
mechanical systems of finite dimensional Hilbert spaces. 
The result is true irrespective of whether the initial state is \emph{pure} or \emph{mixed}.
\end{abstract}
\vskip 2pc] 
\vskip 2pc 
PACs number(s): 03.67.-a, 03.65.Ta
\section{The Relation for Spin - 1/2} 
Consider an ensemble ($N$ copies-$N$ very large) of Spin - 1/2 particles and 
dividing it into 3 equal sub-ensembles. 
Three independent observables are measured and the resulting {\it mixed}
subensembles are put together to form the {\it post-measurement density matrix}. 

Suppose the measured observables are
${\bf S}_{1}$, ${\bf S}_{2}$ and ${\bf S}_{3}$ along three {\it orthonormal}
directions. Let 
$p_{1}$ be the probability for the outcome
$|+\ra_{1}$, $p_{2}$ the probability for
$|+\ra_{2}$ and $p_{3}$ for $|+\ra_{3}$. Further let ${{\bf P}_{+,i}}$ be
the {\it projection operator} for the i-th outcome etc.

Now, the post-measurement density matrices of the individual measurements
are respectively:
\beeq
{\bf \rho}_{i} = p_{i}{\bf P}_{+,i} + (1 - p_{i}){\bf P}_{-,i} 
\eneq

The {\it spectral representation} for $S_i$
\beeq
{\bf S}_i = {1\over 2}({\bf P}_{+,i}-{\bf P}_{-,i})
\eneq
along with the {\it completeness} relation
\beeq
{\bf P}_{+,i}+{\bf P}_{-,i} = {\bf I}
\eneq
yields
\beeq
{\bf \rho}_i = {{\bf I}\over 2}+2<S_i>{\bf S}_i
\eneq
with $<S_i> = 2p_i-1$. An {\it equal} mixture
of these three leads to 
the post-measurement density matrix:
\beeq
{\bf \rho}_{msmt} = {{\bf I}\over 3}+{1\over 3}({\bf I\over 2}+\sum_i 2<S_i>{\bf S}_i)
\eneq
Since this is a complete measurement, the initial density matrix can 
be completely determined and is
\beeq
{\bf \rho}_{ini}= {{\bf I}\over 2} +\sum_i 2<S_i>{\bf S}_i
\eneq
Clearly, there is a linear and universal relation between ${\bf \rho}_{msmt}$ 
and ${\bf \rho}_{ini}$: 
\beeq
\label{msmt}
{\bf \rho}_{msmt} = (1/3) (\bf{I} + {\bf \rho}_{ini})
\eneq
This {\it new relationship} between the pre-and post-measurement
states is the main result of this paper.  

Though the relation $(\ref{msmt})$ was shown to be true for an initial 
{\it pure} state, it is straightforward to see that it holds even when the 
initial state is {\it mixed}. To see this let
$${\bf \rho}_{ini}^{mixed} = \sum_{i} c_{i} {\bf \rho}_{i}^{(0)};~~~~~  (\sum_{i} c_{i} = 1)$$
with ${\bf \rho}_{i}^{(0)}$ all being pure states each of which leads to ${\bf \rho}_{msmt, i}$. Thus we have
$${\bf \rho}_{msmt} = \sum_{i} c_{i} {\bf \rho}_{msmt,i} = (\bf{I} + 
{\bf \rho}_{ini}^{mix})/3$$ 

When the initial state is {\it pure}, the eigenvalues of ${\bf \rho}_{ini}$
are $(1,0)$. It then follows that the eigenvalues of ${\bf \rho}_{msmt}$
are $(2/3,1/3)$ allowing us to write the spectral decomposition
\beeq
{\bf \rho}_{msmt} = {2\over 3}|l\ra\la l|+{1\over 3}|s\ra \la s|
\eneq
where $|l\ra,|s\ra$ are the corresponding eigenstates, Their completeness
\beeq
|l\ra\la l|+|s\ra\la s| = {\bf I}
\eneq
and eqn(\ref{msmt}) lead to the interesting result
\beeq
\label{largest}
{\bf \rho}_{ini} = |l\ra\la l|
\eneq
i.e the eigenstate of ${\bf \rho}_{msmt}$ with the {\it largest} eigenvalue
is the original pure state itself.

\subsection{Measurements along non-orthogonal directions.}
At first it might appear that the relationship eqn(\ref{msmt})is only 
a consequence of 
the measurement being {\it complete}. Now, we show that such a {\t universal} 
relation does not hold 
if the measurements are made along three non-collinear directions which 
also constitute a {\it complete} measurement.
Let $\hat{n}_{1}$, $\hat{n}_{2}$ and $\hat{n}_{3}$ be three unit vectors along which measurements are made such that $\hat{n_{1}}.(\hat{n_{2}} \times \hat{n_{3}}) \neq 0$.
Let $\vec{{\bf S}}.\hat{n}_{i}$ be the three spin-components being measured. 
Let $|+\ra_{\hat{n}_i}$ and $|-\ra_{\hat{n}_i}$ be the eigen-vectors of 
${\vec {\bf S}}.{\hat{n}_i}$ and ${\bf P}(\hat{n}_i,\pm)$ be the corresponding projectors.
As before,
\beeq
{\bf P}(\hat{n}_i,+)+{\bf P}(\hat{n}_i,-) = {\bf I}
\eneq
and
\beeqa
\vec {{\bf S}}.\hat{n}_i &=&{1\over 2}({\bf P}(\hat{n}_i,+)-{\bf P}(\hat{n}_i,-))\nonumber\\
{\bf \rho}_{i} &=& {\bf I}/2 + 2\la \vec{S}.\hat{n}_{i}\ra \vec{{\bf S}}.\hat{n}_i
\eneqa
Hence the post measurement state if the three mixed states are further
mixed with weights $x_i$ ($\sum x_i = 1)$ is:
\beeq
\label{rhoorthog}
{\bf \rho}_{msmt}= {\bf I}/2 + \sum 2x_i \la \vec{S}.\hat{n}_{i}\ra \  \vec{{\bf S}}. 
\hat{n}_{i}
\eneq
Now,
$$\rho_{ini} = {\bf I}/2 + 2\la \vec{S} \ra. \vec{{\bf S}}$$
and $\la\vec{S}\ra$ must be expressed in terms of the observed components :
Let,
\beeq
\label{matrix}
\la \vec{S}\ra = \sum c_{i} \hat{n}_{i};~~~~~~ 
\la \vec{S}.\hat{n}_{i}\ra = \sum_j \hat{n}_i.\hat{n}_j c_j
\eneq
The matrix $\hat{n}_i.\hat{n}_j$ is {\em invertible}. Inverting the matrix 
eqn(\ref{matrix}) one gets 
\beeq
c_{i} = \sum_{j} d_{ij} \la\vec{S}.\hat{n}_j \ra;~~~~
\la \vec{S} \ra = \sum_{ij} d_{ij} \la \vec{S}.\hat{n}_j\ra \hat{n}_{i}
\eneq 
where $d_{ij}$ are functions of $\{ \hat{n}_{i}. \hat{n}_{j} \}$ only and have no dependance on $\la \vec{S}.\hat{n}_{i} \ra$.
Finally 
\beeq
{\bf \rho}_{ini} = \textbf{I}/2 + \sum_{ij} 2d_{ij} \la\vec{S}.\hat{n}_j\ra \vec{\bf {S}}. \hat{n}_{i}
\eneq
Consider the average of $\vec{\bf {S}}.\hat{n}_i$ in ${\bf \rho}_{msmt}$; on
using eqn(\ref{rhoorthog}) one gets
\beeq
\label{exptn}
\la \vec{S}. \hat{n}_{i}\ra_{\rho_{msmt}} =\sum_j x_j \hat{n}_i.\hat{n}_j~  \la \vec{S}. \hat{n}_{j}\ra_{\rho_{ini}} 
\eneq
where $\hat{n}_{i}.\hat{n}_{j} \neq 0$. 
Also, $x_i\neq 0$ and are {\em independent} of the $\la \vec{S}.\hat{n}_{i} 
\ra$. This immediately leads to a contradiction because \emph{if} 
we assumed  $\rho_{msmt} = \alpha \rho_{ini} + \beta I$ 
(where $\alpha$ and $\beta$ are constants) then,
$$\la \vec{S}.\hat{n}_{i} \ra_{\rho_{msmt}} = \alpha \la  \vec{S}.\hat{n}_{i} \ra_{\rho_{ini}}$$
which is clearly not of the form of (\ref{exptn})

\section{Generalization to arbitrary ${\cal H}$:}
We now state our result as a theorem: \linebreak
\textbf{Theorem:} \emph{For a quantum system with Hilbert space of 
dimensionality $N$ (complex), {\em only} the post-measurement state resulting from 
complete measurements made in {\it Mutually Unbiased Basis} (MUB) 
and a {\it Mutually Unbiased Mixture}(MUM) have a simple linear and universal 
relation with the pre-measurement or initial state: 
${\bf \rho}_{msmt}$ $=$ $(I + {\bf \rho}_{initial})/(N + 1)$}.

\textbf{Proof :} We first show that it is sufficient to have MUB with MUM for
the result to hold. Let ${\cal H}$ be the Hilbert Space of the considered quantum system and let
$\dim({\cal H}) = N$.
Every basis of this space has $N$ vectors each of $N$ components. A density 
operator describing such a quantum system is an $N \times N$ {\it Hermitean} 
matrix with {\em unit trace}
and having $(N^2 - 1)$ independent real parameters in general (but fewer 
for pure states). Any measurement will yield $(N - 1)$ independent real numbers.
Therefore, $\frac{N^2 - 1}{N-1}$ $=$ $N + 1$ independent observables are 
required to be measured to make a complete measurement. However, certain 
observable sets are more useful  than others and in particular observables 
whose eigenstates form the so-called Mutually Unbiased Bases \cite{woot} 
have been shown to yield what are called 'optimal' measurements which 
minimise the error matrix. These have also been shown to have important 
information theoretic properties \cite{kra}. MUB bases satisfy the 
property $|\la v_{m}|w_{n}\ra|$ $=$ $1/\sqrt{N}$ where $v_{m}$ and $w_{n}$ 
are vectors belonging to \emph{different} basis of an $N$-dimensional space. 
It has also been shown that there are exactly $(N + 1)$ bases that are MUB. 

Let $X^{(\alpha)}$ be the set of observables whose eigenstates are 
$|k^{(\alpha)}\ra$.
These observables are {\it independent} so that their measurement would constitute
a {\it complete} measurement. Therefore the set label $\alpha$ takes on $N+1$ 
values
and the state label $k$ takes $N$ values. Let the eigenstate corresponding
to the outcome $j$ when $X^{(\alpha)}$ is measured be denoted $|j^{(\alpha}\ra$
and ${\bf P}^{(\alpha)}_j$ be the associated projector. Further, let $p^{(\alpha)}_j$ be the corresponding probability. So one has
$$
{\bf P}^{(\alpha)}_j = |j^{(\alpha)}\ra\la j^{(\alpha)}|;
~~{\bf P}^{(\alpha)}_i{\bf P}^{(\alpha)}_j = \delta_{ij}{\bf P}^{(\alpha)}_i;
~~tr {\bf P}^{(\alpha)}_j=1$$
The corresponding post-measurement density matrix is given by
\beeq
\label{ABC}
{\bf \rho}^{(\alpha)} = \sum_{i = 1}^{N} p_{i}^{(\alpha)}{\bf P}_{i}^{(\alpha)} 
\eneq
Let the initial state be of the form,
\beeq
\label{ini}
{\bf \rho}_{ini} =  \sum_{\alpha = 1}^{N + 1} \sum_{i = 1}^{N} c_{i}^{(\alpha)} {\bf P}_{i}^{(\alpha)}
\eneq
where $c_{i}^{(\alpha)}$ are  real parameters(there are only $N^2-1$ independent ones).
The unit trace condition leads to
\beeq
\label{trace}
tr ({\bf \rho}_{ini}) = \sum_{\alpha = 1}^{N + 1} \sum_{i = 1}^{N} c_{i}^{(\alpha)} = 1
\eneq
Consider the expectation value of the operator ${\bf P}_{k}^{(\alpha)}$ in the 
initial state $\rho_{ini}$ 
\beeqa
\label{cintop}
p^{(\alpha)}_k &=& tr {\bf \rho}_{ini} {\bf P}_k^{(\alpha)}= tr~ \sum_{\beta = 1}^{N+1}\sum_{i=1}^N c_i^{(\beta)}{\bf P}_i^{(\beta)}{\bf P}_k^{(\alpha)}\nonumber\\
&=& c_k^{(\alpha)}+\sum_{\beta\ne\alpha=1}^{N+1}\sum_{i=1}^N c_i^{(\beta)}
tr {\bf P}_k^{(\alpha)}{\bf P}_i^{(\beta)}\nonumber\\
&=& c_k^{(\alpha)}+{1\over N}(1-\sum_i c_i^{(\alpha)})
\eneqa
where we used eqn(\ref{trace}) along with the MUB property $tr~~ {\bf P}_i^
{(\alpha)}{\bf P}_j^{(\beta\ne\alpha)} = {1\over N}$ for {\em all} $(i,j,\alpha,\beta\ne\alpha)$.
Taking ${\bf \rho}_{msmt}$ to be an \emph{Unbiased Mixture} of all $\rho^{(\alpha)}$'s one gets, 
\beeq
\label{MM}
{\bf \rho}_{msmt} = \frac{1}{N + 1} \sum_{\alpha = 1}^{N + 1} \rho^{(\alpha)}
= {1\over N+1}\sum_{i\alpha}p_i^{(\alpha)}{\bf P}_i^{(\alpha)}
\eneq
Substituting for $p_i^{(\alpha)}$ from eqn(\ref{cintop}) one gets
\beeqa
\label{result}
{\bf \rho}_{msmt} &=&{1\over N+1}\sum_{i\alpha}{\bf P}_i^{(\alpha)}[c_i^{(\alpha)}+{1\over N}(1-\sum_m c_m^{(\alpha)})]\nonumber\\
&=&{{\bf I}\over N+1} +{1\over N+1}{\bf \rho}_{ini}
\eneqa
where we used eqn(\ref{ini}) as well as $\sum_i {\bf P}_i^{(\alpha)}={\bf 1}$ and
$\sum_\alpha (1-\sum_m c_m^{(\alpha)}) = N$. In proving the latter use has
been made of eqn(\ref{trace}). This completes the proof. 

However, in this proof assumptions were  made in $(\ref{MM})$ of the 
post-measurement density matrix being an {\bf Unbiased Mixture} of all the 
$\rho^{(\alpha)}$'s; the set of observables $X^{(\alpha)}$ were also
assumed to be such that their eigenfunctions form MUB's. We shall now show 
that it is unnecessary to make these assumptions and in fact it can be
proved that {\it Mutually Unbiased Bases} as well as {\it Mutually
Unbiased Mixtures} are the {\bf only ones} that can lead to the result in
question.

Let us now consider the expectation value of ${\bf P}_k^{(\alpha)}$
in ${\bf \rho}_{msmt}$ with {\em unequal weights} instead. It then follows that
\beeqa
\label{msmt1}
tr~{\bf \rho}_{msmt}{\bf P}_k^{(\alpha)}&=&\sum_{j\beta}~c^{(\beta)}p_j^{(\beta)}tr~{\bf P}_k^{(\alpha)}{\bf P}_j^{(\beta)}\nonumber\\
&=&c^{(\alpha)}p_k^{(\alpha)}+\sum_{j,\beta\ne\alpha}c^{(\beta)}p_j^{(\beta)}tr~{\bf P}_k^{(\alpha)}{\bf P}_j^{(\beta)}
\eneqa
On the other hand, if a relation of the type
\beeq
{\bf \rho}_{msmt} = {\lambda\over N}{\bf I}+(1-\lambda){\bf \rho}_{ini}
\eneq
were to hold with $\lambda = const$ we should also have
\beeq
\label{msmt2}
tr~{\bf \rho}_{msmt}{\bf P}_k^{(\alpha)}= {\lambda\over N}+(1-\lambda)p_k^{(\alpha)}
\eneq
Now eqns(\ref{msmt1},\ref{msmt2}) subject to the constraint $\sum_i p_i^{(\alpha)}=1$ must be true for {\em arbitrary} $p_i^{(\alpha)}$. Using
Lagrange multipliers $\mu^{(\alpha)}$ for the constraints we get
\beeqa
0&=& \sum_\alpha \mu^{(\alpha)}(\sum_i p_i^{(\alpha)}-1)+c^{(\alpha)}p_k^{(\alpha)}\nonumber\\
&+&\sum_{j,\beta\ne\alpha}c^{(\beta)}p_j^{(\beta)}tr~{\bf P}_k^{(\alpha)}{\bf P}_j^{(\beta)}\nonumber\\
&-&{\lambda\over N}-(1-\lambda)p_k^{(\alpha)}
\eneqa
Now coefficients of $p_j^{(\gamma)}$ can be set equal to zero; this way we
get the conditions
\beeqa
0&=&1-\lambda-c^{(\alpha)}-\mu^{(\alpha)}\nonumber\\
0&=& \mu^{(\alpha)}\nonumber\\
0&=& c^{(\beta\ne\alpha)}tr~ {\bf P}_k^{(\alpha)}{\bf P}_j^{(\beta\ne\alpha)}+\mu^{(\beta\ne\alpha)}
\eneqa
which can be solved to yield
\beeqa 
c^{(\alpha)} &=&1-\lambda\nonumber\\
tr~ {\bf P}_k^{(\alpha)}{\bf P}_j^{(\beta\ne\alpha)} &=& -{\mu^{(\beta)}\over c^{(\beta)}}
\eneqa
The second of these equations says that $ tr~ {\bf P}_k^{(\alpha)}{\bf P}_j^{(\beta\ne\alpha)}$ is {\em independent} of $(j,k)$. But on remembering $\sum_j {\bf P}_j^{(\gamma)}={\bf I}$ one sees that this is possible iff 
\beeq
tr~ {\bf P}_k^{(\alpha)}{\bf P}_j^{(\beta\ne\alpha)} = {1\over N}
\eneq
In other words, the bases spanned by the eigenstates of $X^{(\alpha)}$ {\em are}
{\em MUB}. Furthermore, there is nothing special about the value of $\alpha$
used above which means $c^{(\alpha)}$ are the same for {\em all} values
of $\alpha$, and each equal to ${1\over N+1}$, which proves that the mixture has to be {\em mutually unbiased}.
Finally, these considerations fix $\lambda={N\over N+1}$. 
\linebreak

We end our calculations by formulating the criterion for getting the special relationship in terms of the mixture weights and the bases alone without any
reference to the initial state or the post-measurement state. For this,
let $\rho_{ini}$ be represented, in
one element of the set of $N+1$ basis vectors, say, $|i^{(0)}\ra$, be
\beeq
\rho_{ini} = \sum_{k,l}~~|k^{(0)}\ra \la l^{(0)}|~~\rho^{ini}_{kl}
\eneq
Denoting the probability of getting the eigenvalue labelled by $j$ upon 
a measurement of $X^{(\alpha)}$ by $p^{(\alpha)}_j$ one has
\beeqa
\label{palphaj}
p^{(\alpha)}_j&=& tr~\rho^{ini}~{\bf P}_j^{(\alpha)}\nonumber\\
              &=&\sum_{kl}\rho^{ini}_{kl}\la j^{(\alpha)}|k^{(0)}\ra~\la l^{(0)}
	         |j^{(\alpha)}\ra\nonumber\\
  &=&\sum_{kl}\rho^{ini}_{kl}C^{(\alpha)}_{jl}C^{(\alpha)*}_{jk}		 
\eneqa
where  
\beeq
C^{(\alpha)}_{ij} = \la j^{(0)}|i^{(\alpha)}\ra
\eneq
The mixed state that results after this measurement is:
\beeq
\rho^{(\alpha)} = \sum_j~p^{(\alpha)}_j{\bf P}_j^{(\alpha)}
\eneq
Let these be mixed with weights $c^{(\alpha)}$ with $\sum_\alpha c^{(\alpha)}=1
$. The resulting post-measurement density matrix is given by
\beeq
\rho_{msmt} = \sum_{i\alpha}c^{(\alpha)}p^{(\alpha)}_i {\bf P}_i^{(\alpha)}
\eneq
We express this in the same basis $|i^{(0)}\ra$ in which we had expressed
$\rho_{ini}$
\beeqa
\rho^{msmt}_{kl}&=& \sum_{i\alpha} c^{(\alpha)} p^{(\alpha)}_i \la k^{(0)}|
                    i^{(\alpha)}\ra \la i^{(\alpha)}|l^{(0)}\ra\nonumber\\
&=& \sum_{i\alpha}c^{(\alpha)}p^{(\alpha)}_i C^{(\alpha)}_{ik} C^{(\alpha)*}_{il}    
\eneqa
Substituting the expression for $p^{(\alpha)}_i$ from eqn(\ref{palphaj}) one has
\beeqa
\rho^{msmt}_{kl}&=&\sum_{i\alpha} c^{(\alpha)}C^{(\alpha)}_{ik} C^{(\alpha)*}_{il}
\sum_{pq}\rho^{ini}_{pq}C^{(\alpha)}_{ip} C^{(\alpha)*}_{iq}\nonumber\\
&=& \sum_{pq} d_{kl,pq}\rho^{ini}_{pq}
\eneqa
where
\beeq
\label{dklpq}
d_{kl,pq} = \sum_{i\alpha} c^{(\alpha)} C^{(\alpha)}_{ik} C^{(\alpha)*}_{il} 
C^{(\alpha)}_{ip} C^{(\alpha)*}_{iq}
\eneq
Since we are looking for a relationship of the type
$$
\rho_{msmt} = {\lambda\over N}{\bf I} + (1-\lambda)\rho_{ini}
$$
we must have
\beeq
\label{dklpq2}
d_{kl,pq} = {\lambda\over N}\delta_{kl}\delta_{pq}+(1-\lambda)\delta_{kp}\delta_{lq}
\eneq
Equating eqn(\ref{dklpq}) to eqn(\ref{dklpq2}) we get the required criterion
to be
\beeq
\sum_{i\alpha} c^{(\alpha)} C^{(\alpha)}_{ik} C^{(\alpha)*}_{il} 
C^{(\alpha)}_{ip} C^{(\alpha)*}_{iq}
= {\lambda\over N}\delta_{kl}\delta_{pq}+(1-\lambda)\delta_{kp}\delta_{lq}
\eneq
This can be taken as an alternate definition of MUB.

Now we establish the equivalent of eqn(\ref{largest}) for the generalised
case. Again note that ${\bf \rho}_{msmt}$ has one eigenvalue ${2\over N+1}$
and $N-1$ smaller eigenvalues ${1\over N+1}$ leading to the spectral
decomposition
\beeq
{\bf \rho}_{msmt} = {2\over N+1}|l\ra\la l|+{1\over N+1}\sum_{i=1}^{N-1}|s_i\ra\la s_i|
\eneq
Now using eqn(\ref{result}) one finds, as before,
\beeq
{\bf \rho}_{ini} = |l\ra\la l|
\eneq

We finally comment on a very similar looking relation first derived by
Audretsch et al \cite{audretsch}. Their eqn(20) looks remarkably like our
eqn(\ref{result}) but these equations and the respective contexts are 
very different. Firstly their eqn(20) is only for {\em fideleties} while
our result is for {\em density matrices}. Secondly the {\it pre} and {\it post}
quantities appear {\em oppositely} to our equation. Finally, their results
are derived in the context of so called {\it generalised measurements}
while our considerations are in the context of {\it projective measurements}.
As generalised measurements are more general than projective measurements,
the precise connection between these equations is interesting to pursue.
\section{Acknowledgements}
CD thanks Prof. Ajay Patwardhan of St. Xavier's College, Mumbai, for his 
invaluable   encouragement, support and guidance over the years as also 
for his commitment towards his students. CD also thanks The Institute of 
Mathematical Sciences for its hospitality and support through a Summer 
Students' Fellowship. 

\end{document}